# High performance artificial visual system with plasmon-enhanced 2D material neural network

Tian Zhang, Xin Guo, Pan Wang, Linjun Li*, Limin Tong

**State Key Laboratory of Modern Optical Instrumentation, College of Optical Science and Engineering, Zhejiang University, Hangzhou, China**

**Intelligent Optics and Photonics Research Center, Jiaxing Institute Zhejiang University, Jiaxing, China**

Corresponding author

Correspondence to: lilinjun@zju.edu.cn; orcid.org/0000-0002-2734-0414

## Abstract

Artificial visual systems (AVS) have gained tremendous momentum because of its huge potential in areas such as autonomous vehicles and robotics as part of artificial intelligence (AI) in recent years. However, current machine visual systems composed of complex circuits based on complementary metal oxide semiconductor (CMOS) platform usually contains photosensor array, format conversion, memory and processing module. The large amount of redundant data shuttling between each unit, resulting in large latency and high power consumption, which greatly limits the performance of the AVS. Here, we demonstrate an AVS based on a new design concept, which consists of hardware devices connected in an artificial neural network (ANN) that can simultaneously sense, pre-process and recognize optical images without latency. The Ag nanograting and the two-dimensional (2D) heterostructure integrated plasmonic phototransistor array (PPTA) constitute the hardware ANN, and its synaptic weight is determined by the adjustable regularized photoresponsivity matrix. The eye-inspired pre-processing function of the device under photoelectric synergy ensures the considerable improvement of the efficiency and accuracy of subsequent image recognition. The comprehensive performance of the proof-of-concept device demonstrates great potential for machine vision applications in terms of large dynamic range (180 dB), high speed (500 ns) and ultralow energy consumption per spike ($2.4 \times 10^{-17}$ J).

## Introduction

The human visual system is mainly composed of the eyes and the visual cortex of the brain[1,2]. The retina of the eye is normally used to capture external optical information and perform first-stage image pre-processing[3-5]. The regulated visual signals are transmitted to the neural network of the visual center for final processing and recognition[6,7]. Accordingly, a variety of bio-inspired AVS for emulating certain functions of the human eye and neural network image processing have emerged that are used to perform typical image processing functionalities, which include image-contrast enhancement[1,2,8,9], noise suppression[10,11], visual adaptation[5,12], detection and recognition[13-18], and auto-encoding[19]. However, for current AVS, a hardware solution with both the pre-processing function of the human retina and the image recognition capability of the visual cortex has not been reported, especially in time-critical applications[18,19]. There is a high demand to develop multifunctional electronic devices to meet the challenges of next generation machine vision. Additionally, developing low-power and high-efficiency AVS has become a major research focus, where the most critical issue to be addressed is the efficient conversion of optical images into electrical digital signals.

Plasmonic energy conversion has been considered as a promising alternative to drive a wide range of physical and chemical processes[20]. This emerging method is based on the generation of hot electrons with energy distribution deviating substantially from equilibrium Fermi-Dirac distribution in plasmonic nanostructures after light absorption through non-radiative electromagnetic decay of surface plasmons[21-25]. while the 2D semiconductor itselfhas excellent optoelectronic properties[28-30] such as ultrafast response[31,32], external tunability[19,33] and large photothermoelectric effect[34]plasmonics can further enable strong light-matter interactions in 2D materials[26,27],. 2D materials technology has by now achieved a sufficiently high level of maturity for integration with conventional complex electronic systems[35,36,38]. Herein, we present a PPTA constructed of nanogratings and 2D heterostructures, which constitutes an ANN that integrates simultaneous sensing, pre-processing and image recognition functions. The plasmonic phototransistor (PPT) takes advantage of the strong coupling of photonic and electronic resonances in an elaborately designed device, in which hot electrons are injected efficiently into the floating gate and produce a large photoelectric effect, to simulate the response of the human retina to optical color information. Moreover, the electrical dynamic modulation of the gate electrode can effectively enlarge the dynamic range of the device for image pre-processing functions (image contrast enhancement). Further real-time image recognition is realized by training the network through varying the drain-source voltage to set the photoresponsivity value of each pixel individually. As a result, the AVS integrated with image pre-processing and ANN can effectively improve the image quality, and increase the efficiency and the accuracy of image recognition.

## Results

Figure 1a illustrates the schematic structure of a 2D PPT, which consists of a 2D $MoS_2$/Ag nanograting integrated structure on the left and a 2D $MoS_2$/h-BN/$WSe_2$ heterostructure on the right. The left part of the device mimics the sensing and pre-processing functions of the human retina for color information (Extended Data Fig.1a) using light-excited waveguide-plasmon polaritons (WPPs)[37] and electrical modulation of the gate electrode, respectively (see Fig. 2 for more details of the mechanism). The photocurrent signal processed in the first stage can be passed to the floating gate on the right side of the device to induce the channel current, which is similar to that visual information can be transmitted through the optical nerve to each neuron in the visual center via synaptic interconnection (Extended Data Figs. 1a, b). The photoresponsivity (synaptic weight) of the device is modulated by changing the drain-source voltage to emulate the regulation of neurotransmitter release between biological synapses (Extended Data Fig. 1c). To avoid unnecessary direct photocurrents in the channel, the right side is covered by the $Al_2O_3$/Au layer. Interconnecting each 2D PPT (subpixel) in the form of an ANN constitutes an AVS with image sensing, pre-processing and recognition functions (Fig. 1b). It contains $N$ pixels, which form the imaging array, and each pixel is divided into $M$ subpixels. The circuit connections of $M$ subpixels and $N$ pixels are presented in Figs. 1c,d, respectively. Each subpixel delivers a photocurrent of $I_{mn} = R_{mn}P_n$ under illumination, where $R_{mn}$ is the regularized photoresponsivity of the subpixel and $P_n$ denotes the optical power at the *n*th pixel. $n = 1, 2, ..., N$ and $m = 1, 2, ..., M$ denote the pixel and subpixel indices, respectively.

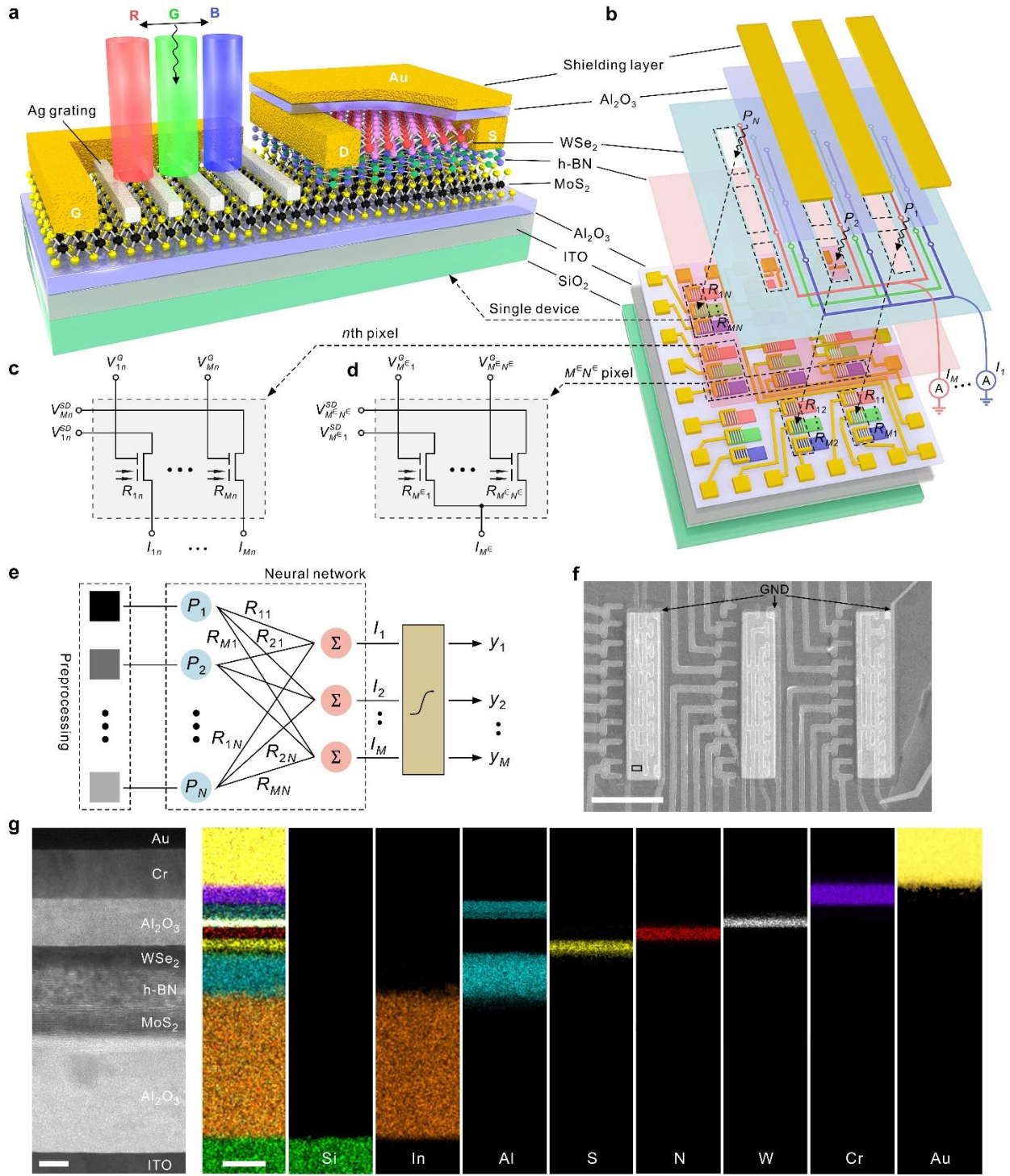

**Fig.1|AVS-inspired 2D ANN PPTA. a**, Schematic of a 2D PPT. **b**, Disassembled diagram of the 2D ANN PPTA. The current induced by subpixels of the same color in WSe$_2$ channel layer is connected in parallel by wires of the same color to generate an output current $I_M$. **c, d,** Circuit diagram of the *n*th pixel (c) and $M^\in N^\in$ subpixels (d) in the array, where $M^\in$ is a subset of *M*, representing a certain number of subpixels among *M* subpixels with the same *M* indice, and $N^\in$ is a subset of *N*, representing a certain number of pixels among *N* pixels. **e**, Illustration of an AVS based on the 2D PPT for image pre-processing and an ANN for image recognition. **f**, Scanning electron microscopy (SEM) image of the PPTA. Scale bar, 20 μm. GND, ground electrode. **g**, High-resolution scanning transmission electron microscope image captured from the black box in (f) and energy dispersive X-ray spectroscopy mapping. Scales bar are 8 (left) and 40 nm (right), respectively.

The schematic of a classifier is provided in Fig. 1e. The array is operated as a single-layer perceptron using pre-processed visual information as the input layer. Here, we chose the softmax function $\phi_m(I) = e^{I_m \xi} / \sum_{k=1}^{M} e^{I_k \xi}$ as the nonlinear activation function to generate the neuron output off-chip, where $\xi = 10^{11} \text{A}^{-1}$ is a scaling factor. In one type of ANN representing a supervised learning algorithm, in order to facilitate the classification of images **P** into different categories **y**, we chose a binary code encoding, where each of the three letters corresponds to an output code. Following the elaborated design concept of the 2D PPTA, we fabricated the actual device as shown in Fig. 1f. The sample fabrication process is provided in Extended Data Fig. 2 (for details, see Methods). This device consists of 27 subpixels ($N \times M = 27$), of which every 9 subpixels were arranged to form a $3 \times 3$ imaging array ($N = 9$) with a subpixel size of about $17 \times 5$ μm$^2$. A schematic of the entire circuit connections of the array is presented in Extended Data Fig. 3. Summing all photocurrents generated by 9 PPTs with the same subpixel index *m* according to Kirchhoff's law, the output $I_m$ is expressed as

$$I_m = \sum_{n=1}^{N} I_{mn} = \sum_{n=1}^{N} R_{mn} P_n \qquad (1)$$

Figure 1g shows the high-resolution scanning transmission electron microscope and energy dispersive X-ray spectroscopy element mapping characterizations of a single subpixel in the black box in Fig. 1f, indicating a clean heterostructure interface.

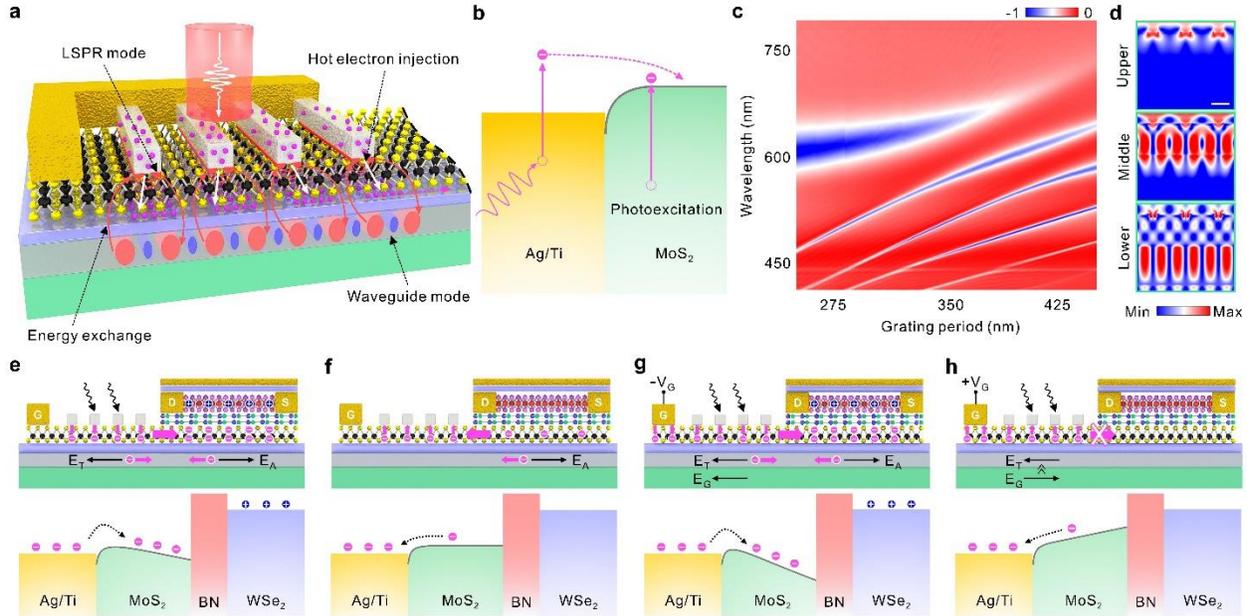

**Fig.2 | Schematic illustrations of the mechanisms of 2D PPT. a**, A diagrammatic model proposed to describe the whole physical process of the strong coupling between the LSPR mode and the waveguide mode and its relaxation. **b**, Simplified band diagram illustrating the hot electron injection process taking place at the Ag-MoS$_2$ interface. In addition to receiving hot electrons emitted from the Ag nanograting, MoS$_2$ itself can also generate a small amount of electrons after receiving light. **c**, The simulated transmittance spectrum of the Ag nanograting-ITO waveguide integrated structure dependent on the grating period, showing the classic Rabi splitting. **d**, The calculated electric field distribution at the 320 nm grating period corresponding to each branch at red (R), green (G) and blue (B) wavelengths in the strong coupling regime. Scale bar, 180 nm. **e-h**, Charge-flow illustrations and schematic band diagrams at different operation modes: light on (**e**), light off (**f**), light on and apply $-V_G$ (**g**), light on and apply $+V_G$ (**h**). The blue balls denote the holes, the magenta balls denote the electrons, and the magenta arrows indicate the flow direction of the electrons. $E_T$ represents the thermoelectric potential, $E_A$ represents the accumulated

potential, and $E_G$ represents the gate potential. The black arrows represent the direction of each potential. The black dotted arrow represents the direction of the electron transition.

In order to understand the mechanism of 2D PPT, we present a scenario for elaboration below. As shown in Fig. 2a, following light absorption and localized surface plasmon resonance (LSPR) excitation in the Ag nanograting, the electromagnetic resonance can be damped radiatively by re-emission of photons, or non-radiatively through transferring the energy to hot electrons via Landau damping[20,21]. In the subsequent hot electron injection[24] (Fig. 2b), hot electrons with momentum within the escape cone[23] can be rapidly emitted into MoS$_2$ through ohmic contacts during the relaxation time[22,27]. At the same time, 2D MoS$_2$ itself also produces a fraction of energetic hot electrons after absorbing light energy, although the effect of this fraction is minimal (see Extended Data Fig. 4a-c). Figure 2c shows the simulated normalized transmittance mapping of the grating period from 250 to 450 nm in the visible region (for details, see Methods), where Rabi splitting can be clearly observed as a distinguishing characteristic of the strong coupling. It is worth mentioning that the upper, middle and lower three hybrid branches are caused by the coupling of the symmetric and antisymmetric modes in the waveguide with the LSPR mode, respectively, and the bottom two branches are caused by the presence of the mode in the quartz substrate, which is independent of the strong coupling modes. We choose the three eigenenergies corresponding to the red (632 nm), green (535 nm) and blue light (469 nm) when the grating period is 320 nm as the eigenvalues of the three-coupled oscillator model to analyze the strong coupling of this structure. The obtained Rabi splitting ($\Omega \approx 680$ meV) is satisfied with the strong coupling criterion between these three oscillators, that is, $\Omega > \mathbf{W} \cdot \sum_{i=Pl,\ Sym,\ Asym} \mathbf{P}^i \gamma_i$, where $\mathbf{W} = (W_{Upper},\ W_{Middle},\ W_{Lower})$ are the weight of each hybrid branch, $\mathbf{P}^i = \left(P^i_{Upper},\ P^i_{Middle},\ P^i_{Lower}\right)$ represents the proportion of uncoupled states in each branch, and $\gamma_i$ represents the linewidth of each uncoupled mode. (for details, see Methods). The electric field distribution corresponding to the eigenenergy of different branches at the period 320nm is provided in Fig. 2d. It can be clearly found that the coupling between LSPR mode and waveguide mode leads to energy exchange. The above mechanism suggests that the 2D PPT can respond to optical color information, and it is also the first time that the splitting of three absorption peaks in the visible range has been achieved compared to previous studies[27,37]. Thus, by exploiting the hybrid LSPR and waveguide modes, we realize highly efficient photoelectric conversion, while the limitation on the narrow responding wavelength of LSPR could be surmounted by adjusting the dimension of the Ag nanograting structure.

On the other hand, the hot electrons that can not be emitted from the decay of plasmons can generate enormous heat on the picosecond scale, which leads to a balance between thermoelectric potential $E_T$ (Extended Data Fig. 4d-f) and the accumulated electropotential $E_A$ as shown in Figs. 2e,f[34]. With such mechanism, the device can respond to different luminance (gray scale of image). When the light is turned on and the negative side gate voltage $-V_G$ is applied, the electrons will be more easily transferred from the left side of MoS$_2$ to the right side, as there is an additional gate potential $E_G$ (Fig. 2g). Accordingly, the larger channel current will be induced by the floating gate. Conversely, by applying a positive gate voltage $+V_G$ while the light is turned on, the electrons will be dragged to the left side because of the additional gate potential $E_G$ (Fig. 2h).The holes left on the right side of the floating gate lead to electron doping to the channel, which gives low conductance since WSe$_2$ is a p type semiconductor. The mechanism of the device described in Figs. 2g,h can be used to eliminate the redundant information. Finally, the regulation of the photoresponsivity of a single device can be realized by changing the drain-source voltage, which can be used to train the weights in the ANN formed by interconnected devices.

Having described the design concept of AVS, we next present its feasibility from an experimental perspective. The optical experimental setup is shown in Extended Data Fig. 5a,b and the electrical experimental setup is shown in Extended Data Fig. 6a (for details, see Methods). Here we choose the red light of $\lambda$ = 635 nm, and its power (0-10 μW) is divided into 11 orders. Figure 3a presents the multi-state photocurrents corresponding to different levels of optical power. These photocurrents are graphically visualised as 11 grey levels in the normalized 0-1 interval. By measuring the photocurrent corresponding to three wavelengths of light at the same power P = 10 µW, we can distinguish red (635 nm), green (532 nm) and blue colors (473 nm) when $V_{DS}$ = 0.1 V (Fig. 3b). This is caused by the different absorption rates of the device for the corresponding three wavelengths of light in the strong coupling mechanism, as shown in Fig. 3c. Next, we performed photocurrent-voltage ($I_{PH}$-$V_{DS}$) characteristic measurements under different optical powers (Fig. 3d). It shows a linear dependence of the photocurrent on the voltage over a wide voltage range, which indicates that the device is dominated by ohmic contacts. Then, we extracted photocurrent as a function of optical power under different $V_{DS}$ values (inset in Fig. 3e). An almost symmetrical and adjustable (trainable) linear photoresponsivity between -15 and +15 pA/µW can be obtained by varying the $V_{DS}$ (Fig. 3e). Considering the subsequent ANN training, we plotted the voltage tunable photocurrents corresponding to each grey level, as shown in Fig. 3f. Similar measurements of the optoelectronic characterization of green and blue light and the uniformity of each device are presented in Extended Data Fig. 7a-i. We also performed the photoresponsivity measurement when $V_G$ = -1 V (Extended Data Fig. 8a-i), and the increase of photoresponsivity in the order of magnitude can be applied to image detection and recognition under weak light. Figure 3g-i show the transfer characteristic curves of the PPTs obtained under the illumination with wavelength of 635, 532 and 473 nm and different incident optical power. The dynamic range (DR) is defined by the equation: $DR = 20 \times \log[I_{max}/I_{min}](dB)$, where $I_{max}$ and $I_{min}$ are the photocurrent values corresponding to the maximum and minimum gate voltages, respectively. The calculated effective DR is up to 180 dB, which equals almost the highest value reported up to date[5,11]. Therefore, the characteristic allows us to realize image pre-processing such as contrast enhancement and noise reduction by locally modulating the gate voltage of each pixel.

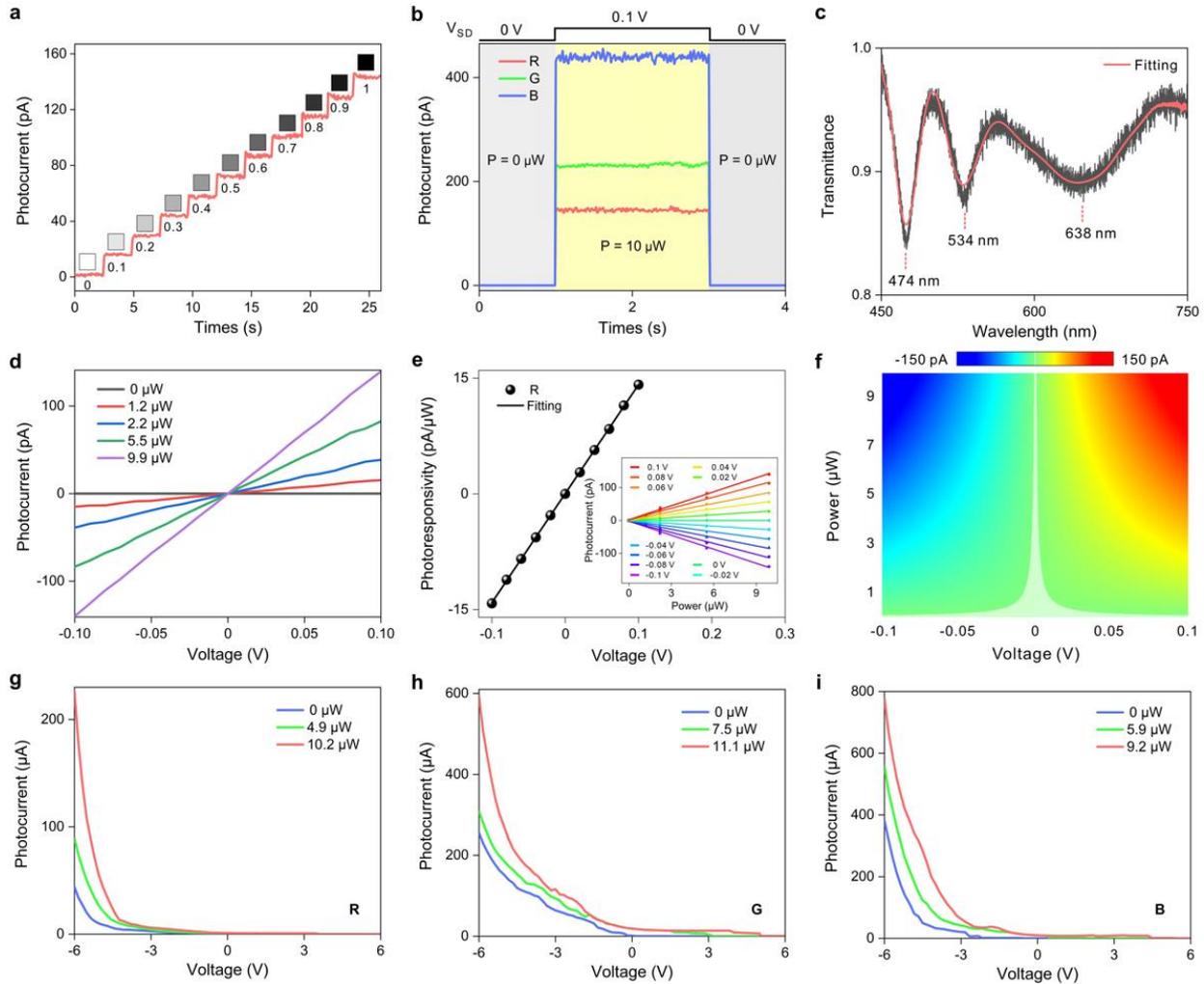

**Fig.3 | Functional implementation of 2D PPT. a**, Multi-state photocurrents corresponding to different levels of optical power (grey levels), where the laser wavelength is 635 nm and the drain-source voltage is 0.1 V. **b**, Photocurrent of different colors of light (R: 635nm, G: 532nm, B: 473nm) under the same measurement conditions, where the power is 10 μW and the drain-source voltage is 0.1 V. **c**, Experimentally measured normalized transmittance spectra of the WPPs structure on the left side of the device. **d**, $I_{PH}$-$V_{DS}$ curves at different optical powers without any applied gate voltage. **e**, Voltage tunability of the regularized photoresponsivity. The inset shows $I_{PH}$ versus P for different $V_{DS}$ values. **f**, The voltage ($V_{DS}$) tunable photocurrent corresponding to each gray scale. **g-i**, The transfer characteristic curves of the devices with red (**g**), green (**h**) and blue (**i**) light measured under different P values at $V_{DS}$=1 V, respectively.

To test the integrated sensing, pre-processing and image recognition functions of the AVS chip, we used it as a classifier to recognize the letters 'z', 'j' and 'u'. For training and testing of the chip, a point-by-point scan is used to project the optical image using the setup shown in Fig. 4a (for details, see Methods). In this example of supervised learning algorithm, cross-entropy is used as the loss/cost function, the weight values were updated by backpropagation of the gradient of the loss function[19]. A detailed flow chart of the whole

AVS including the training algorithm is presented in Extended Data Fig. 6c. Figure 4b illustrates the input image with different Gaussian noise (σ = 0.2, 0.4) added and the pre-processed image (σ = 0.4), which is extracted from the drain-source current $I_D$. After applying gate voltage $V_G$ to the certain pixel (the white pixels in Fig. 4b), the body feature of the letters in the pre-processed image has been enhanced obviously. The complete dataset used for training after pre-processing is given in Extended Data Fig. 9. In Fig. 4c, the accuracy of recognition with and without pre-processing of the images is plotted. For the pre-processed image, it is faster to reach recognition accuracy of 100%. The initial and final responsivities/weights of the classifier are shown in Fig. 4d, and the measured currents and corresponding codes of the target port for each letter are depicted in Fig. 4e. Each code corresponds to a letter, and the corresponding letter is reconstructed through post-processing, as shown in Fig. 4f. To evaluate the overall performance (processing speed and energy consumption) of this network, we also performed time-resolved measurements. The experimental setup is shown in Extended Data Fig. 6b. The trigger/measurement pulse is provided in Extended Data Fig. 10a (see Method for details). The response of a single spike in a single device measured with the assistance of gate voltage is approximately 500 ns (Extended Data Fig. 10b) and the leakage current is shown in Extended Data Fig. 10c. The dissipated energy per spike of the device with such sensitive photoresponse is approximately $2.4 \times 10^{-17}$ J, according to $P = I \times V \times t$[16]. Such a system may hence provide great potential for the development of ultrafast and ultralow power machine vision.

In conclusion, we have presented an AVS composed of PPTA, which integrates multifunctions of sensing, preprocessing and image recognition simultaneously. By performing image pre-processing using this PPT, the image quality is effectively improved, and the efficiency and accuracy of subsequent image recognition is increased. This device exhibits great potential in terms of large dynamic range, ultrafast and ultralow power consumption for machine vision applications.

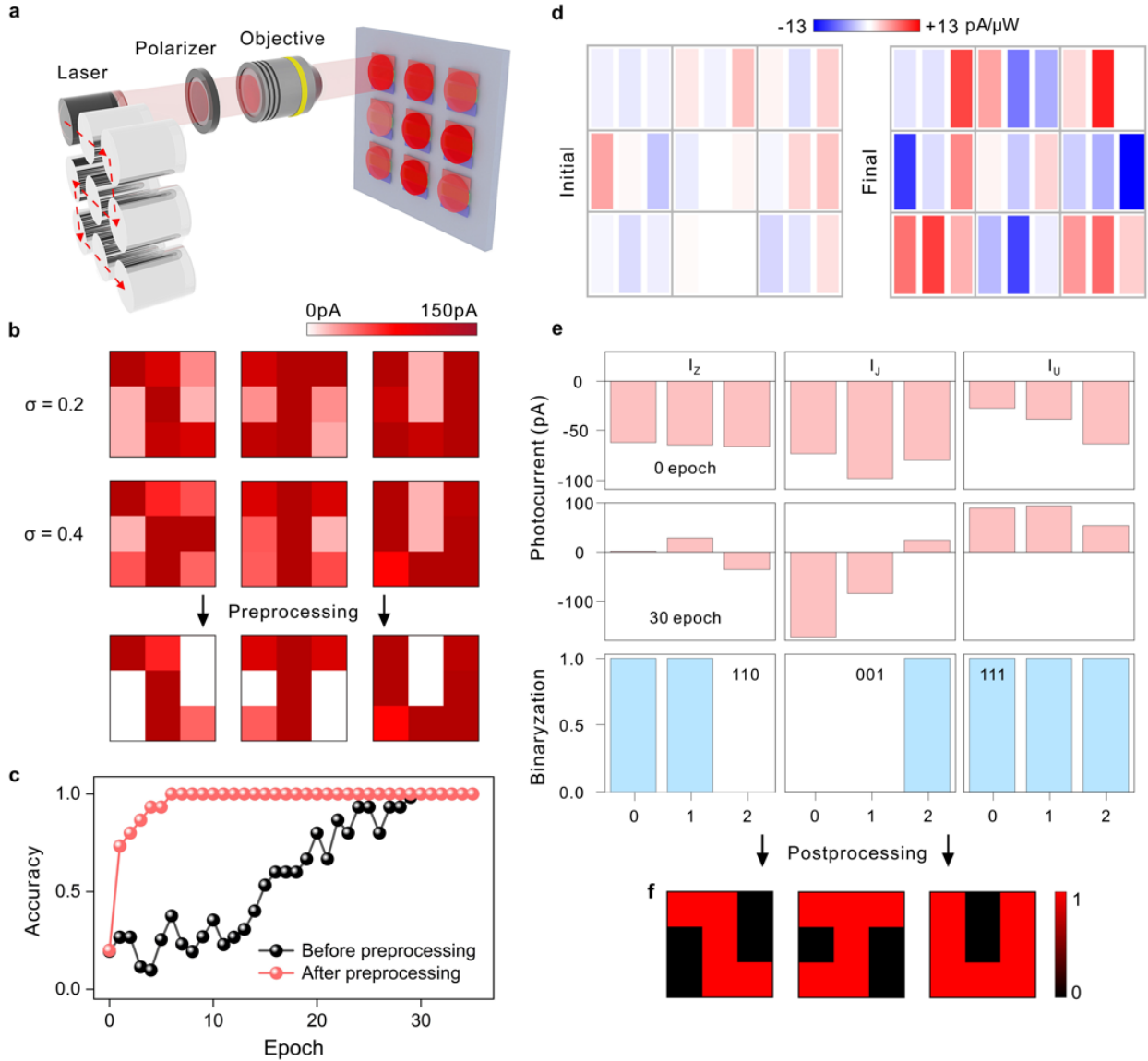

**Fig.4 | AVS operation as a classifier. a**, Schematic illustration of the optical setup for network training/operation. The resulting image is projected onto the photodiode array in a point-by-point scanning manner. **b**, Examples of images with (σ = 0.4) and without (σ = 0.2, 0.4) pre-processing of the device. **c**, Comparison of image recognition rate before and after pre-processing of the device. **d**, Responsivity distributions before (initial) and after (final) training. **e**, The measured three currents corresponding to 'z', 'j' and 'u' target ports, which are converted by the nonlinearity into binary activation codes. In each experiment, the letters 'z', 'j' and 'u' were projected onto the chip separately. **f**, The reconstructed letters after post-processing.

# Methods

## Device fabrication

The fabrication of the chip follows the procedure described in Extended Data Fig. 2. A quartz wafer was used as the original substrate, which was cleaned with acetone, isopropyl alcohol and deionized water, respectively. The cleaned quartz wafer was deposited with a layer of ITO film (~200 nm) using magnetron sputtering(equip? ). Subsequently, an $Al_2O_3$ layer was grew on top of the ITO film by atomic layer deposition (~40nm, Kurt J. Lesker ALD150LX). 2D crystals including $MoS_2$, h-BN and $WSe_2$ flakes were derived from bulk source materials by a mechanical peel-transfer method. For the transfer of $MoS_2$ flake, it was first mechanically exfoliated on a transparent polydimethylsiloxane film and then transferred to the substrate with the help of an optical microscope. To eliminate unnecessary stresses, the transferred 2D $MoS_2$ was annealed in an argon atmosphere. Standard e-beam lithography (EBL, Raith Voyager) and magnetron sputtering were then employed to define the Ti/Ag nanogratings on the produced structures by a lift-off approach. Next, we defined the mask with EBL and carried out reactive ion etching (RIE) with $Ar/SF_6$ plasma to separate the previously transferred $MoS_2$ sheet into 27 pixelsAfterwards, the mask was removed with acetone. 2D h-BN and $WSe_2$ flakes were also transferred to the structure using the same method described above . In order to maximize the absorption of nanogratings, $Ar/SF_6$ plasma was again used to perform RIE towards 2D heterostructure on the mask defined by EBL. The top metal layer (gate electrode and drain-source electrode) was added by another EBL process and Cr/Au (3 nm/15 nm) evaporation. Finally, $Al_2O_3$ (20 nm) and Cr/Au (20 nm/50 nm) layers were deposited on the produced heterostructures by lift-off methods using standard EBL process and magnetron sputtering/thermal evaporation of materials.

## Experimental setup

Schematics of the experimental setup are shown in Extended Data Fig. 5 and Extended Data Fig. 6a, b. Light from a semiconductor laser (635/532/473 nm wavelength) was collimated by a lens before passing through a linear polarizer. The polarization direction of the linear polarizer was mounted perpendicular to the long axis of the Ag nanograting, and the linearly polarized light was projected on the structure in a normal incidence manner. The gray level of each pixel in the optical image was achieved by adjusting the laser power, and then the optical image was projected onto the sample using a microscope objective with a long working distance. A source meter (Keithley, 2400) was used to supply gate voltage to the PPT, and a source meter (Keithley, 2450) was used to supply drain-source voltage to the PPT while measuring the output current. The sample was connected to the source meter via a home-made measurement box and BNC connection cable. For time-resolved measurements, a femtosecond pulsed laser source (BFL-1030-20B, BWT) was used, which was triggered using a lock-in amplifier (Stanford Research Systems, SR830) to emit a single pulse at a wavelength of 515 nm. The 500ns cycle drain-source pulse voltage was provided by an arbitrary waveform generator (Keithley, 3390), and the output current was amplified by a preamplifier (Stanford Research Systems, SR570) and converted into a voltage signal, which was finally recorded by an oscilloscope (Siglent). In addition, all measurements were carried out at room temperature in an air environment.

## Simulation and strong coupling model

The transmittance spectra and electromagnetic field distributions of the structures with strong coupling were simulated using finite-difference time-domain (FDTD) method. The plane wave light source was projected onto the structure with normal incidence in the direction of polarization perpendicular to the long axis of Ag nanogratings. In order to highlight the strong coupling effect, we neglected the effect of 2D materials in our experimental and theoretical simulations. Here, small volumes Ag nanorods with a height of 20 nm were selected to form the grating in order to achieve large photoelectric conversion efficiency by reducing the proportion of radiation damping and increasing the ballistic transport probability[22] and hot electron relaxation time[27].All calculated data were collected while satisfying the steady state energy criteria.

A coupled oscillator model was introduced to analyze the strong coupling behavior of the hybrid architecture under specific parameters. The plasmon of Ag nanogratings, symmetrized photonic mode, and antisymmetrized mode can be assumed as three oscillators. Therefore, the Hamiltonian of this three-coupled system can be written as:

$$\begin{pmatrix} E_{Pl} - i\gamma_{Pl}/2 & g_w & g_s \\ g_w & E_{Asym} - i\gamma_{Asym}/2 & 0 \\ g_s & 0 & E_{Sym} - i\gamma_{Sym}/2 \end{pmatrix} \quad (1)$$

Where $\gamma_{Pl}$, $\gamma_{Asym}$, and $\gamma_{Sym}$ are the linewidths of plasmon, antisymmetrized and symmetrized modes, $E_{Pl}$, $E_{Asym}$, and $E_{Sym}$ are the resonance energies of plasmon, antisymmetrized and symmetrized modes, while $g_w$ and $g_s$ represent plasmon-antisymmetrized mode and plasmon-symmetrized mode interaction constants. In the three-oscillator model, the eigenstates of Hamiltonian correspond to the three hybrid branches. The wave function of each branch from the admixture contribution of plasmon, symmetric mode and antisymmetric mode can be expressed as $|\psi_j\rangle = \alpha_{Pl}^j |Pl\rangle + \alpha_{Sym}^j |Sym\rangle + \alpha_{Asym}^j |Asym\rangle$, where $\alpha_i^j\ (i = Pl, Sym, Asym;\ j = Upper, Middle, Lower)$ denotes Hopfield coefficients. The modular square of the Hopfield coefficient represents the proportion of uncoupled states $\mathbf{P}^i = \left(P_{Upper}^i, P_{Middle}^i, P_{Lower}^i\right)$ in hybrid state. Also, the weight of each hybrid branch $\mathbf{W} = \left(W_{Upper}, W_{Middle}, W_{Lower}\right)$ in this strong coupling regime can be calculated as $W_j = \gamma_j / \sum_j \gamma_j$.

## Data Availability

The data that support the findings of this study are available from the corresponding authors upon reasonable request.

## Acknowledgements

This work was supported by the National Key R&D Program of China (2019YFA0308602), the National Science Foundation of China (general program 12174336 & major program 91950205) and the Natural Science Foundation of Zhejiang Province (LR20A040002). We thank the Micro and Nano Fabrication Centre at Zhejiang University for facility support and W. Wang at the State Key Laboratory of Modern Optical Instrumentation for suggestions on nanofabrication. We appreciate the equipment support provided by the Center of Electron Microscopy of Zhejiang University for the preparation of samples to be characterized, as well as the assistance provided by H. H. Huang from the Center for Micro/Nano Fabrication of Westlake University for sample characterization. We also acknowledge useful comments from Prof. D. Xiang of Frontier Institute of Chip and System, Fudan University.

## Author contributions

L.L. and T.Z. conceived and designed the project. T.Z. designed and built the experimental setup, programmed the machine-learning algorithm, fabricated the ANN PPTA, carried out the material and device characterization. T.Z. and L.L analyzed data and wrote the manuscript. All authors commented on the manuscript.

## Competing interests

The authors declare no competing interests.

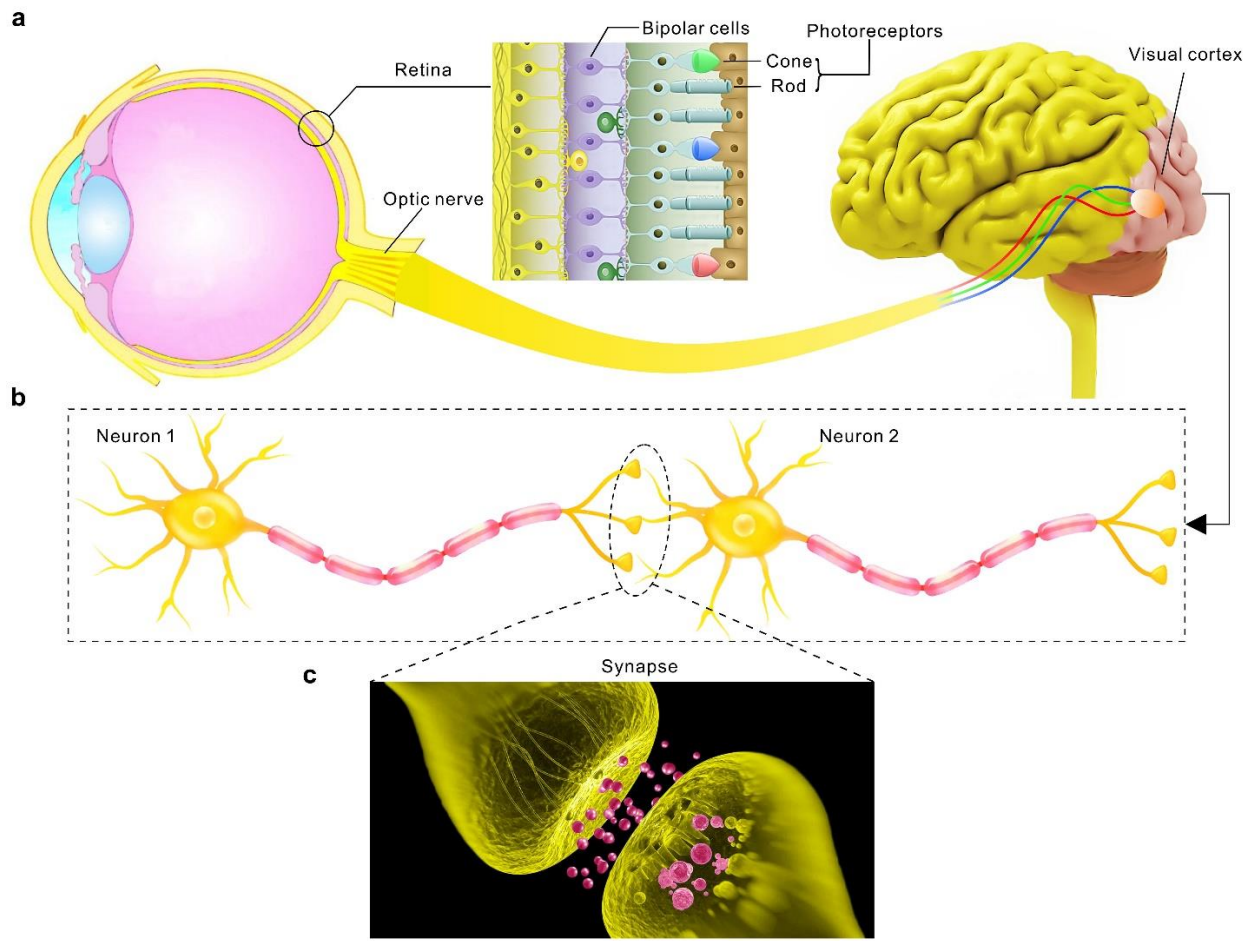

**Extended Data Fig. 1|Schematics of the human visual system. a**, Illustration of an neuromorphic visual system. A biological visual system consisting of the retina (sensing and pre-processing), optic nerve (transducing) and the visual cortex (image recognition). The inset shows a magnified view of the retina organized in a hierarchical way. Photoreceptors (cone and rod cells) are used to sense external color optical information, and the information can be pre-processed through ON and OFF bipolar cells[8]. **b**, Schematic diagram of a neural network formed by synaptic connections between neurons. **c**, Diagram of neurotransmitter transmission between synapses.

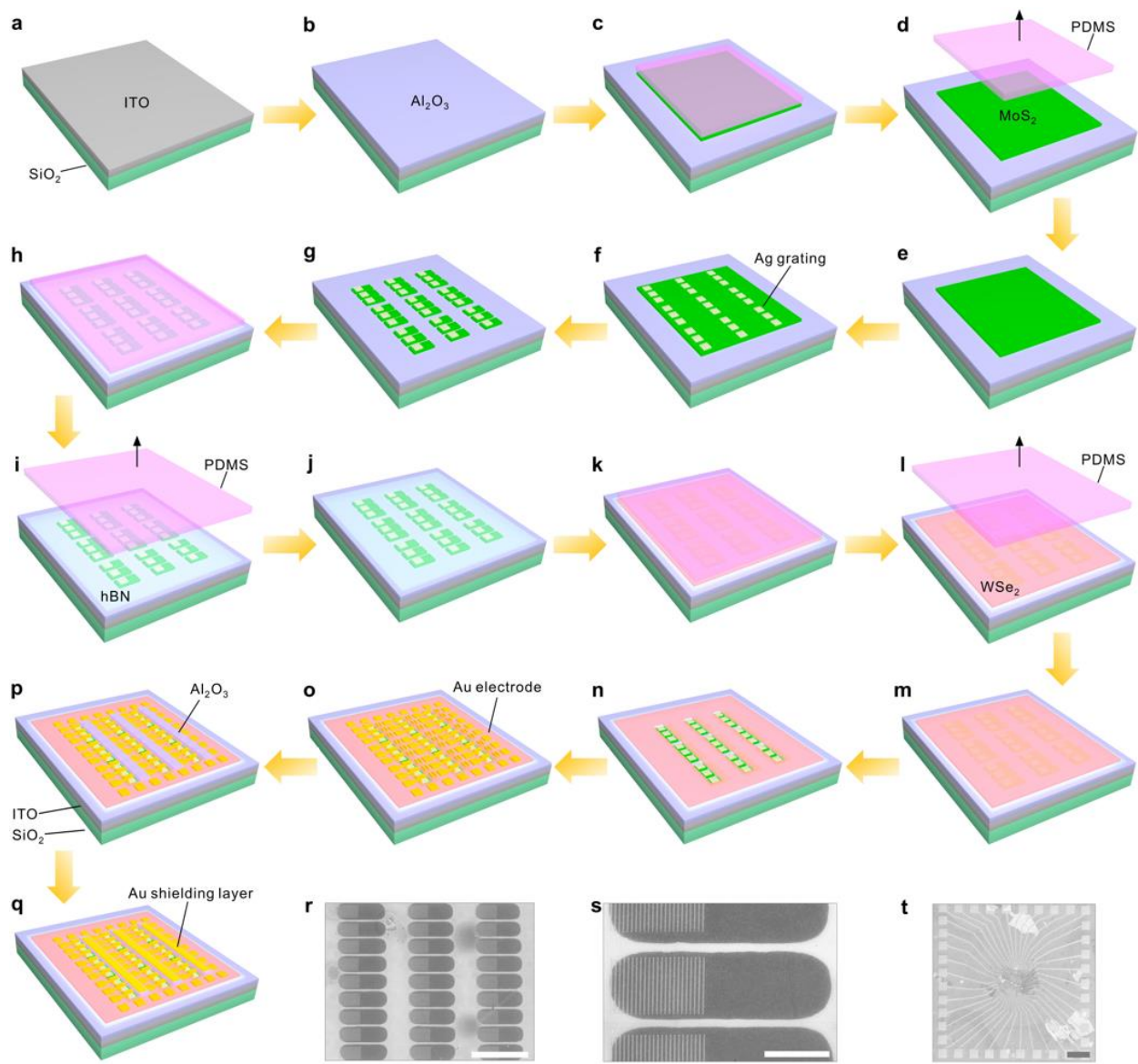

**Extended Data Fig. 2|Schematic fabrication process of the device. a**, Sputtering a layer of indium tin oxide (ITO) on $SiO_2$ substrate by magnetron sputtering. **b**, Growth of a thin film of $Al_2O_3$ using atomic layer deposition technology. **c**, An $MoS_2$ flake is first transferred to a transparent PDMS film by mechanical exfoliation method. **d**, **e**, After the PDMS is lifted off, the $MoS_2$ is left on the substrate just like stamping a pattern with a stamp. **f**, Fabrication of Ag nanograting on $MoS_2$ flake using EBL through overlay process. **g**, RIE of the $MoS_2$ flake using a mask made of EBL. **h**, **i**, **j**, Using the same method, an h-BN flake is transferred on the $MoS_2$ flake. **k**, **l**, **m**, With the same method, a $WSe_2$ flake is transferred on the h-BN/$MoS_2$ flake. **n**, RIE of h-BN/ $WSe_2$ flakes using an EBL mask exposes the Ag nanograting to air. **o**, EBL is used

to define Au/Cr electrodes on the obtained heterostructures. **p**, **q**, Finally, the Al$_2$O$_3$ and Cr/Au layers are sputtered sequentially on the EBL-defined pattern. **r**, SEM view of the sample during the (g) processing step. Scale bar, 20 μm. **s**, Enlarged SEM view of the sample in (r). Scale bar, 5 μm. **t**, Full SEM view of the PPTA. Scale bar, 200 μm.

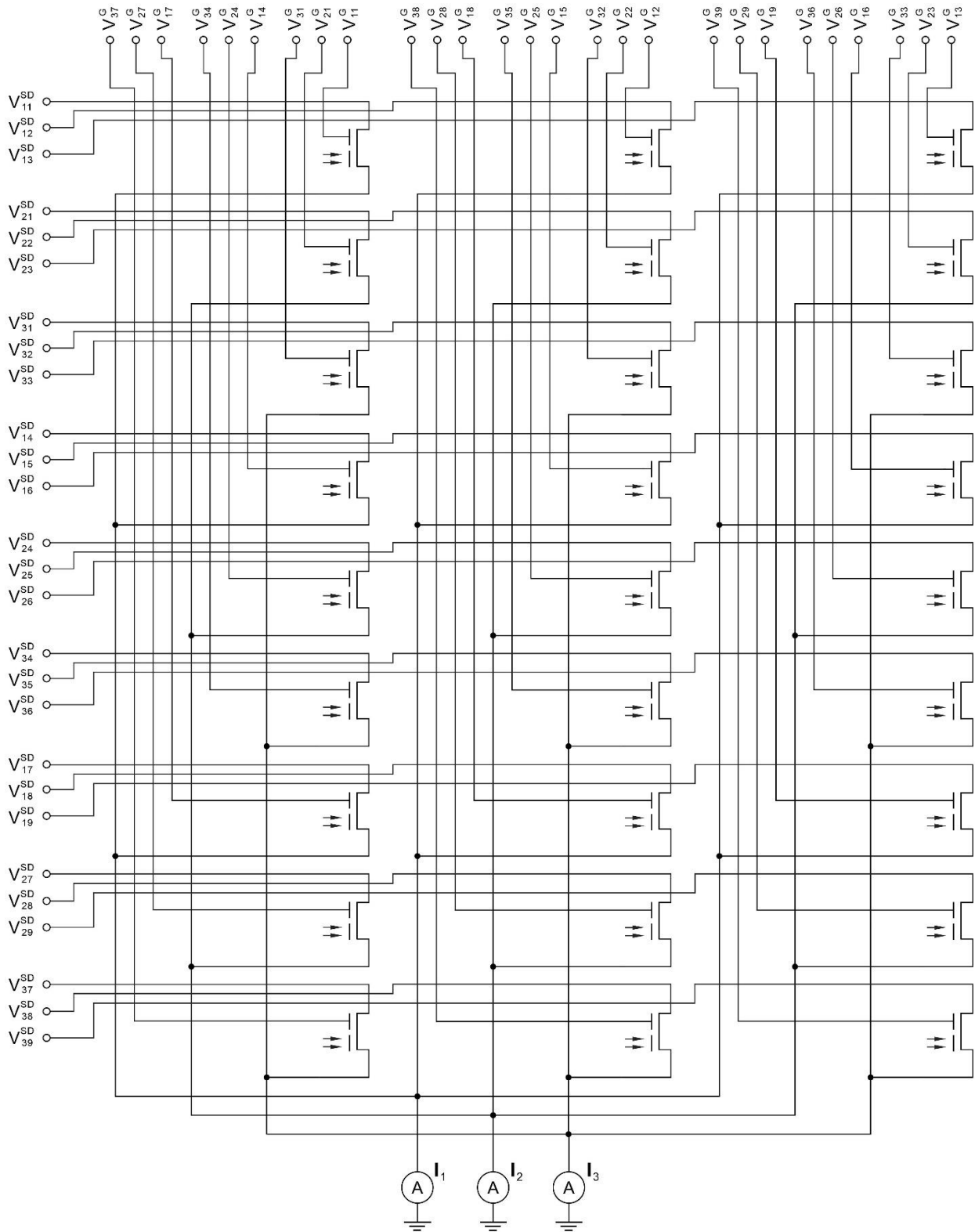

**Extended Data Fig. 3 | Circuit of the ANN PPTA.**

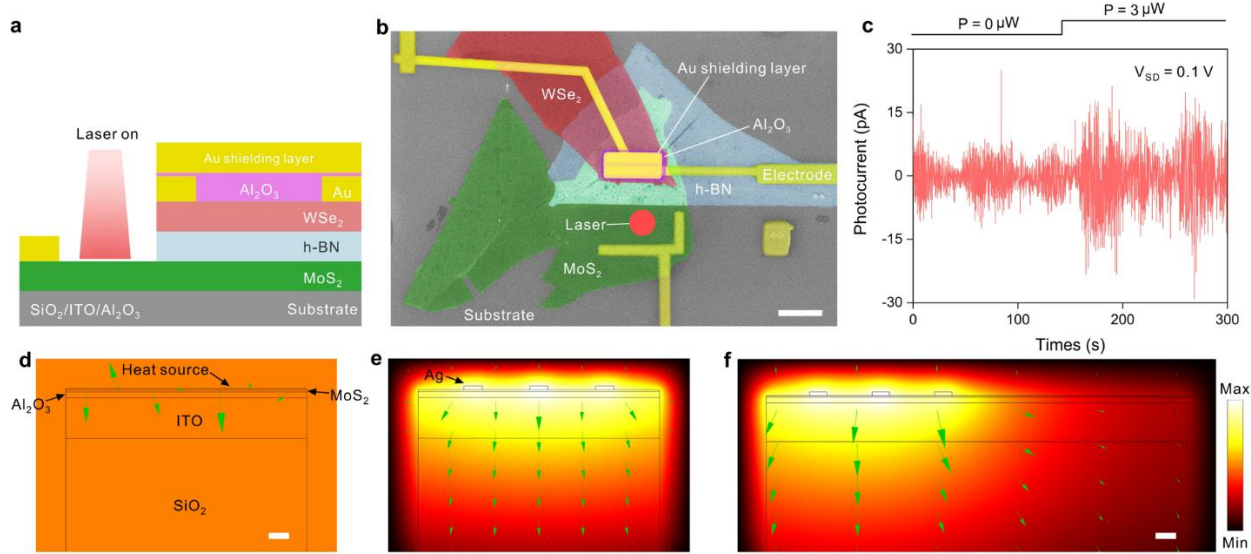

**Extended Data Fig. 4 | Photothermoelectric effect of 2D PPT with and without Ag nanogratings. a**, Schematic of a single 2D PPT without Ag nanogratings. **b**, False-coloured SEM image of a fabricated device without Ag nanograting. The red dot indicates the location of the laser on $MoS_2$. Scale bar, 20 µm. **c**, When the laser is turned on and off, only subtle changes of photocurrent can be detected in the $WSe_2$ channel, indicating that very few electrons in the $MoS_2$ excited by the device without Ag nanograting can transfer to the $MoS_2$ floating gate below the $WSe_2$ channel through the photothermoelectric effect. **d-e**, Simulated temperature distribution of the left part of the phototransistor without (d) and with Ag nanograting (e). The colored images indicate the temperature distribution and the green arrows indicate the heat flux. The heat source (heat power density) is written as $Q = \frac{1}{2}\omega \text{Im}(\varepsilon_r)|\mathbf{E}|^2$, where $\omega$ is angular frequency of the light, $\varepsilon_r$ is the relative permittivity of silver, and E is electric field calculated in Fig. 2d. In the absence of the right part of the device, the heat generated by plasmon decay is mainly transmitted downward, resulting in an increase in the temperature of $MoS_2$. Scale bar, 100 nm. **f**, The simulated temperature distribution of the $MoS_2$ floating gate layer of the device extending to the right. The heat generated by the plasmon decay is transmitted to the lower right, but the heat is still mainly concentrated in the Ag nanograting region, resulting in a thermoelectric potential between the left and right sides of the $MoS_2$ floating gate layer. Scale bar, 100 nm.

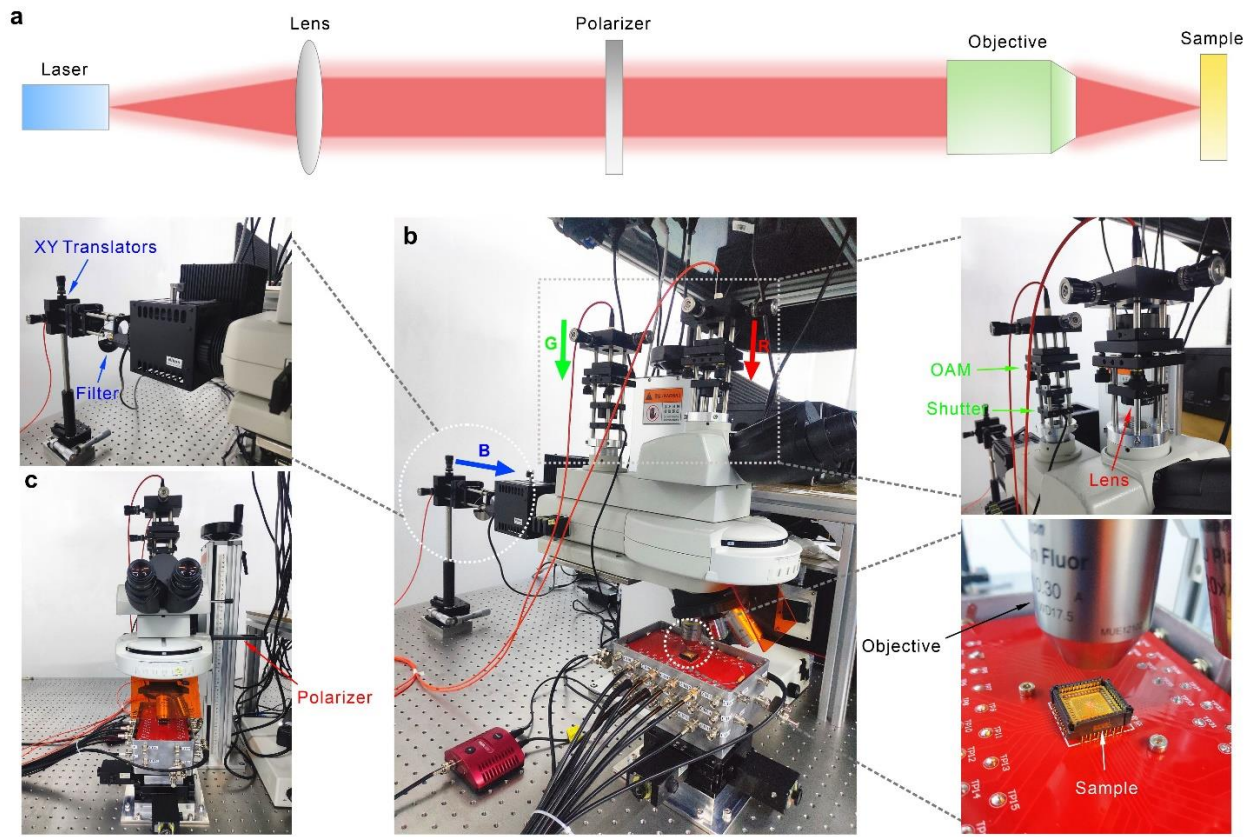

**Extended Data Fig. 5|Optical setup. a**, Schematic illustration of the optical setup. The polarizer enables that the linearly polarized light (TM polarization) generated by the incident laser is perpendicular to the Ag nanograting. **b**, Photograph of the optical setup. OAM: Optical Adjustable Mount. **c,** Front view photo of the optical setup.

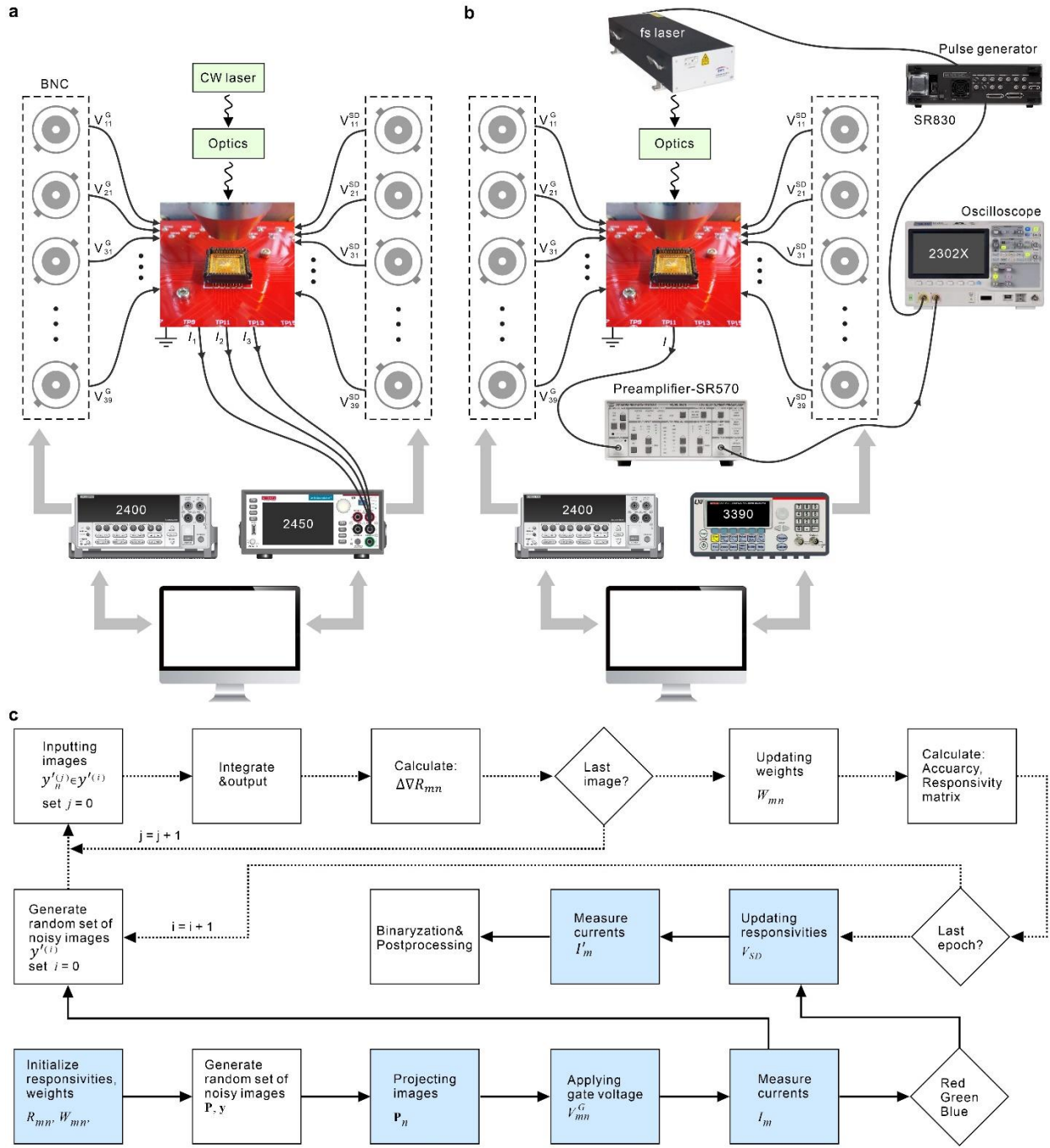

**Extended Data Fig. 6|Experimental setup. a**, Experimental setup of an AVS used as a classifier. CW, continuous wave. Optical setup is shown in Extended Data Fig. 5. **b**, Experimental setup for time-resolved measurements. **c**, Flow chart of AVS used for image processing. The blue shaded boxes indicate the ANN PPTA in action.

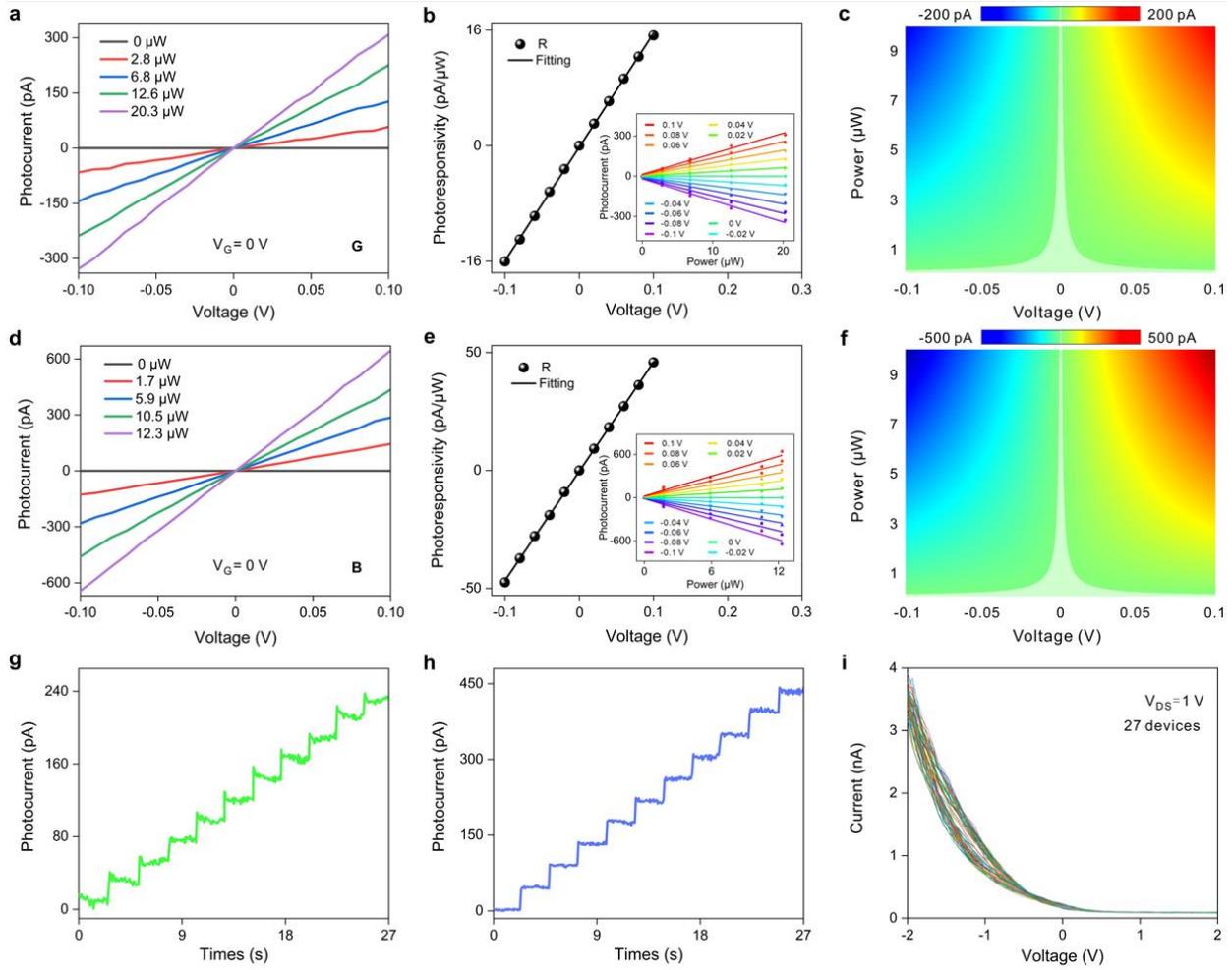

**Extended Data Fig. 7 | Measurement of optoelectronic characteristics of 2D PPT. a**, $I_{PH}$-$V_{DS}$ curves at different green light powers without any applied gate voltage. **b**, Voltage tunability of the regularized photoresponsivity extracted from (**a**). The inset shows $I_{PH}$ versus P for different $V_{DS}$ values. **c**, The voltage ($V_{DS}$) tunable photocurrent (green light) corresponding to each gray scale. **d**, $I_{PH}$-$V_{DS}$ curves at different blue light powers without any applied gate voltage. **e**, Voltage tunability of the regularized photoresponsivity extracted from (**d**). The inset shows $I_{PH}$ versus P for different $V_{DS}$ values. **f**, The voltage ($V_{DS}$) tunable photocurrent (blue light) corresponding to each gray scale. **g-h**. The multi-state photocurrents corresponding to different levels of optical power (gray levels), where the laser wavelengths are 532 nm (**g**) and 473 nm (**h**), respectively, and the drain-source voltage is 0.1 V. **i**, The transfer characteristic curves of 27 devices measured under dark conditions at $V_{DS}$ = 1V.

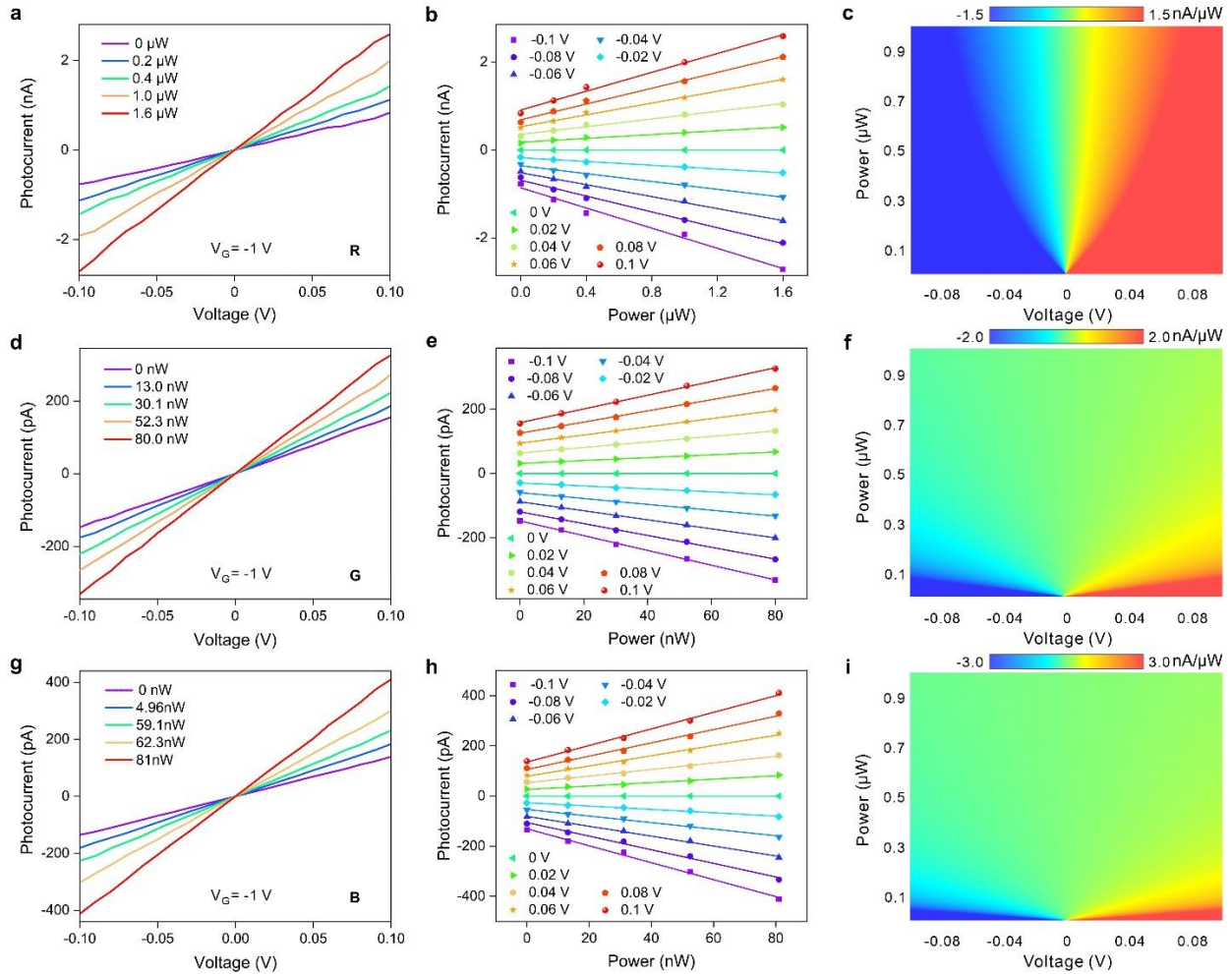

**Extended Data Fig. 8|Implementation of PPT photoresponsivity. a**, $I_{PH}$-$V_{DS}$ curves at different red light powers with applied gate voltage. **b**, The relationship between $I_{PH}$ and P under different $V_{DS}$ values extracted from (**a**). **c**, The voltage ($V_{DS}$) tunable red light regularized photoresponsivity corresponding to each gray scale. **d**, $I_{PH}$-$V_{DS}$ curves at different green light powers with applied gate voltage. **e**, The relationship between $I_{PH}$ and P under different $V_{DS}$ values extracted from (**d**). **f**, The voltage ($V_{DS}$) tunable green light regularized photoresponsivity corresponding to each gray scale. **g**, $I_{PH}$-$V_{DS}$ curves at different blue light powers with applied gate voltage. **h**, The relationship between $I_{PH}$ and P under different $V_{DS}$ values extracted from (**g**). **i**, The voltage ($V_{DS}$) tunable blue light regularized photoresponsivity corresponding to each gray scale.

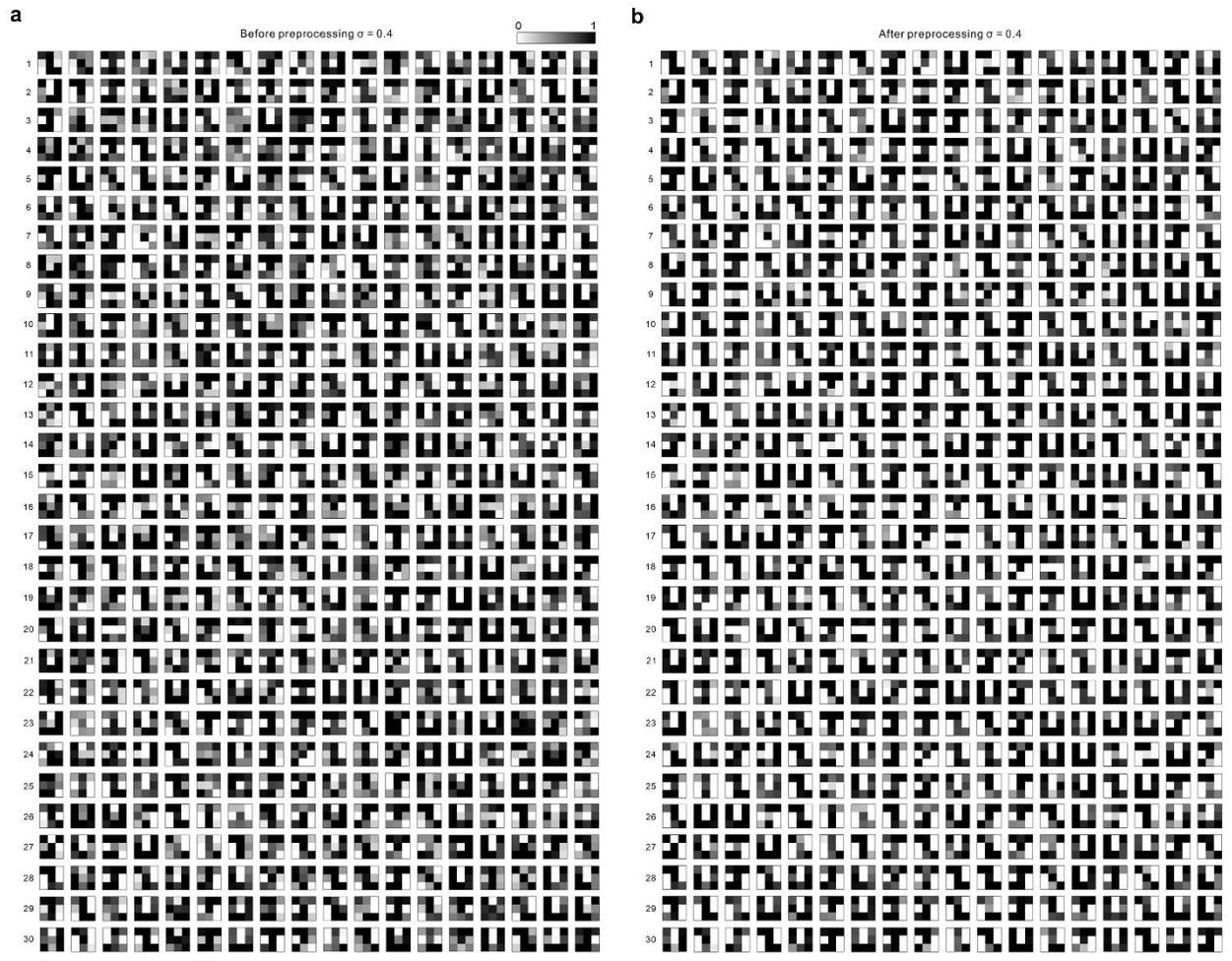

**Extended Data Fig. 9|Training datasets. a**, **b**, Dataset with noise level σ = 0.4 for classifier training with (**a**) and without (**b**) preprocessing.

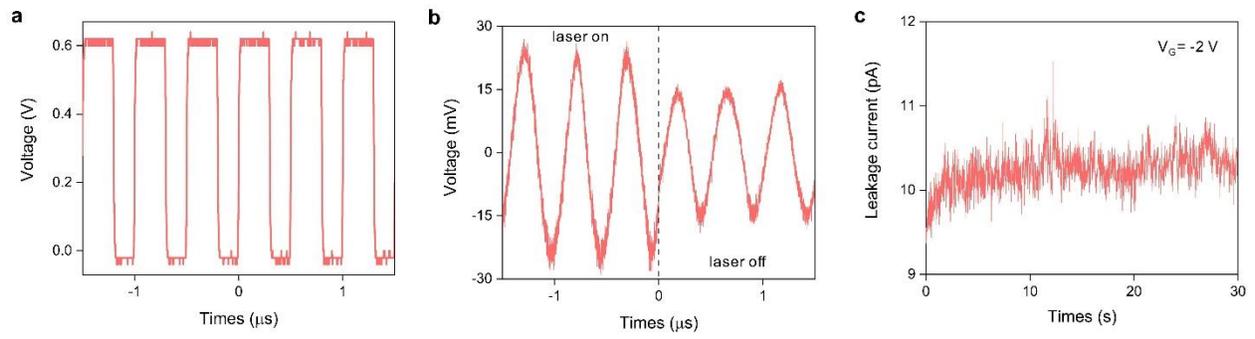

**Extended Data Fig. 10 | Time-resolved measurements. a**, Electrical pulse for synchronously triggering laser and drain-source current measurement. **b**, The drain-source voltage signal when the laser is turned on or off after preamplifier conversion. **c**, Leakage current when side gate voltage is applied to assist measurement.